\documentclass[twocolumn,showpacs,preprintnumbers,prb]{revtex4}
\usepackage{dcolumn}
\usepackage{amssymb}
\usepackage{amsmath}
\usepackage{bm}
\usepackage{graphicx}
\usepackage{float}
\usepackage{subfigure}
\usepackage{epstopdf}
\usepackage{float}

\usepackage{color}

\newcommand{\ket}[1]{| #1 \rangle} % for Dirac bras
\begin{document}

\title{Global entanglement and quantum phase transitions in the transverse XY Heisenberg chain}

\author{Roya Radgohar}
\author{Afshin Montakhab}\email{montakhab@shirazu.ac.ir}

\affiliation{Department of Physics, College of Sciences, Shiraz
University, Shiraz 71946-84795, Iran}

 \pacs{03.67.Mn,03.65.Ud,64.70.Tg,05.50.+q}
\date{\today}

\begin{abstract}
We provide a study of various quantum phase transitions occurring
in the  XY Heisenberg chain in a transverse magnetic field using
the Meyer-Wallach measure of (global) entanglement. We obtain
analytic expression of the measure for finite-size systems, and
show that it can be used to obtain critical exponents via
finite-size scaling with great accuracy for the Ising universality
class. We also calculate an analytic expression for the isotropic
(XX) model and show that global entanglement can precisely
identify the level-crossing points. The critical exponent for the
isotropic transition is obtained exactly from an analytic
expression for global entanglement in the thermodynamic limit.
Next, the general behavior of the measure is calculated in the
thermodynamic limit considering the important role of symmetries
for this limit. The so-called oscillatory transition in the
ferromagnetic regime can only be characterized by the
thermodynamic limit where global entanglement is shown to be zero
on the transition curve. Finally, the anisotropic transition is
explored where it is shown that global entanglement exhibits an
interesting behavior in the finite size limit.  In the
thermodynamic limit, we show that global entanglement shows a
cusp-singularity across the Ising and anisotropic transition,
while showing non-analytic behavior at the XX multi-critical
point. It is concluded that global entanglement can be used to
identify all the rich structure of the ground state Heisenberg
chain.
\end{abstract}

\maketitle

\section{Introduction}\label{Introduction}
Quantum phase transition (QPT) occurs as a result of a sudden
change in the ground state as a system's parameter (e.g.~external
field) is slowly changed [\onlinecite{ref:SQPT}]. Quantum
fluctuations, instead of thermal fluctuations, drive such
transitions, i.e $T\approx 0$. This sudden change is accompanied
with novel behavior on the macroscopic level. QPT has attracted
intense attention in the field of condensed matter physics. The
prominent examples are quantum Hall systems
[\onlinecite{ref:SGCS}], superconductor-insulator transitions
[\onlinecite{ref:HLG}] and heavy-fermion compounds
[\onlinecite{ref:PC}]. It is not unexpected that such interesting
quantum systems should be able to be characterized using tools of
quantum information theory [\onlinecite{ref:NC}]. Consequently, it
has been shown that quantum information measures such as
concurrence [\onlinecite{ref:OAFF02}], entanglement
density [\onlinecite{ref:TWM}], fidelity [\onlinecite{ref:CWHW}],
geometric phases [\onlinecite{ref:Zhu}], quantum discord
[\onlinecite{ref:RD}] and quantum macroscopicity
[\onlinecite{ref:AK}] provide valuable insight into the nature of
QPT. It has been shown that such measures can signal the ensuing
QPT (i.e. critical point) as well as providing scaling behavior
which leads to evaluation of critical exponents. Furthermore,
since entanglement can be used as a resource for quantum
technology, QPT can provide a fertile playground as criticality
implies highly correlated systems which implies maximal
entanglement.
\par
Therefore, entanglement as a function of control parameter, as
well as its scaling behavior, are key issues when studying quantum
critical phenomena.  While early studies focused on bi-partite
measures of entanglement
[\onlinecite{ref:OAFF02},\onlinecite{ref:GTLC},\onlinecite{ref:GLL},\onlinecite{ref:VPM}], it has recently
been argued that a better characterization is provided by
multi-partite measures of entanglement
[\onlinecite{ref:TWM},\onlinecite{ref:montakhab}].
This is particularly important since criticality is achieved by
long-range correlations in short-range interacting systems.  Such
long-range correlations are much better characterized by global
measures. We therefore propose to study the archetypical XY
Heisenberg chain of spin-$1/2$ model using the Meyer-Wallach
[\onlinecite{ref:MW}] measure of global entanglement. The XY chain
in the presence of transverse field plays a central role in
condensed matter theory (e.g. quantum Hall effect
[\onlinecite{ref:SRFB}]), as well as being a good candidate to
connect small quantum processors in quantum computers
[\onlinecite{ref:zheng}] and to transmit information between long
distance sites in quantum communication protocols
[\onlinecite{ref:SB,ref:latorre}].

While being relatively simple, the model exhibits a rich phase
diagram, including a quantum Ising critical transition line, an
anisotropic transition line, and the intersection of these two
lines which provides a unique (XX) critical point. The model also
exhibits a (classical) transition in the ferromagnetic regime
known as the oscillating transition. Most studies have focused on
the Ising critical line which belongs to Ising universality class.
Here, taking advantage of analytic solutions of the model, we
provide expressions for global entanglement. We study all the
above transitions using finite size as well as infinite size
(thermodynamic) limit. We show that global entanglement is capable
of providing important characterizations for each transition
considered, including scaling behavior, level-crossing, and
critical exponents. Our results provide further evidence that a
multi-partite measure of entanglement could act as a thermodynamic
parameter in quantum many-body systems.

\par
The paper is organized as follows: In the next section, we discuss
the XY model and its phase diagram. We also provide a brief
introduction to global entanglement measure which we use
throughout our study. In section III, we first provide an
analytical expression of global entanglement for the XY model and
extract the correlation length exponent by applying finite-size
scaling for the Ising transition. We then calculate the measure
for the XX model, determining the level-crossing points as well as
obtaining the critical exponent. Next, global entanglement
behavior in the thermodynamic limit and the (classical)
oscillating line are considered in section IV. In section V, we
consider the behavior of global entanglement near the anisotropic
phase transition line. We close by summarizing our results and
providing some commentary.

\section{Model and Measure}\label{sec:model}

\begin{figure}
  \center
   \includegraphics[width=0.4\textwidth]{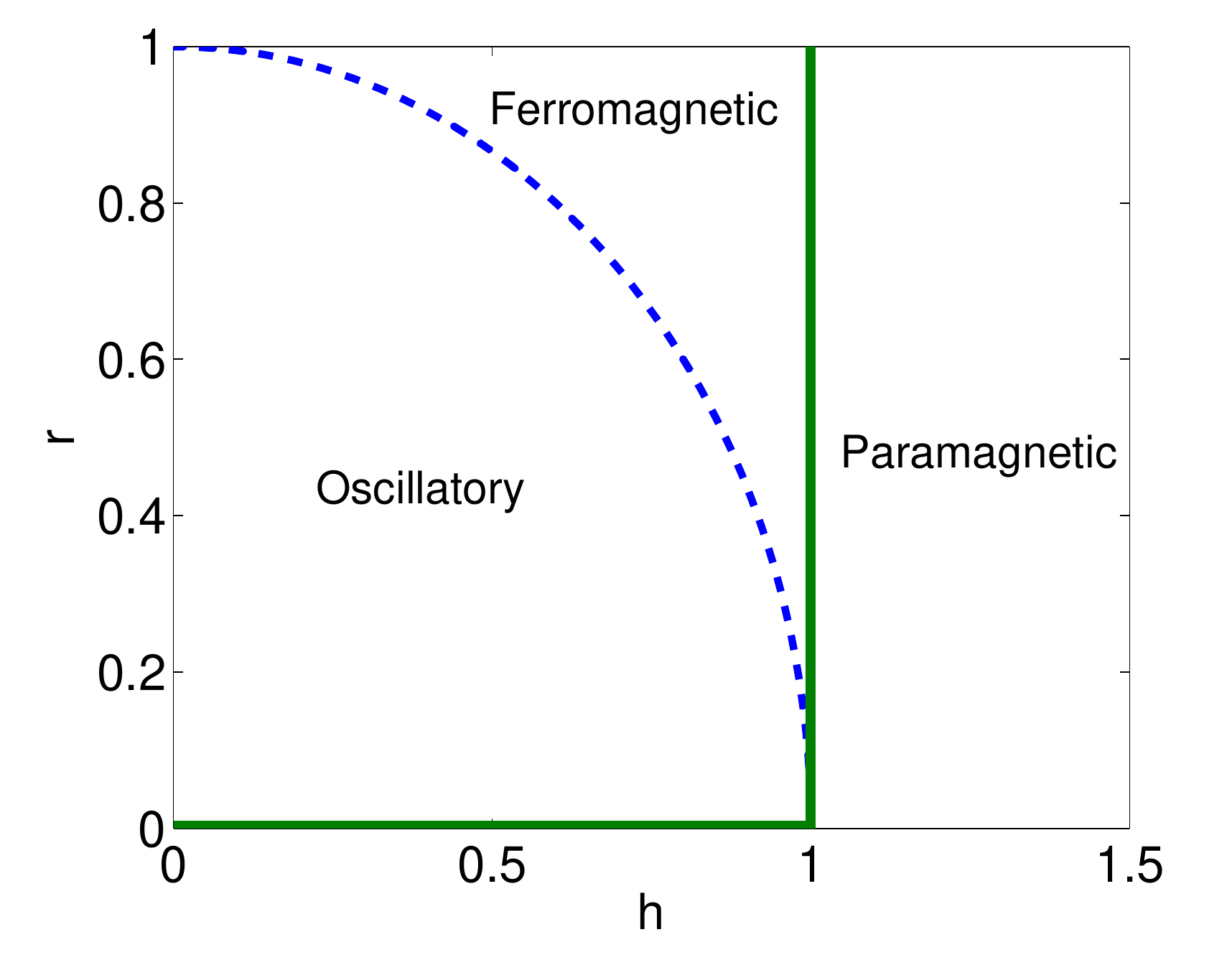}
  \caption{\label{fig:phasediagram}Phase diagram of the 1D XY model.
  Green solid lines $h=1$ (Ising) and $r=0$ (anisotropic) are critical
  lines, their intersection is the multi-critical point.
  Blue dashed curve separates the oscillatory part of the ferromagnetic phase.}
\end{figure}

The system under consideration is a family of models consisting of
$N$ spin-1/2 (qubits) arranged in a chain interacting through
nearest-neighbor coupling and transverse external field in the
z-direction. The Hamiltonian of the system is given by:
\begin{eqnarray}\label{hamiltonian}
  H= \sum_{i=0}^{N-1} -J\Big(\frac{1+r}{2} \sigma_i^x \sigma_{i+1}^x + \frac{1-r}{2} \sigma_i^y \sigma_{i+1}^y\Big)-h\sigma_i^z
\end{eqnarray}
where $\sigma_{i}^{\mu}(\mu=x,y,z)$ are the Pauli matrices, $J$ is
the exchange coupling ($J=1$ in this paper), $h$ is the strength
of magnetic field and $r$ measures anisotropy degree of spin-spin
interactions in the $x-y$ plane which typically varies from $0$
(isotropic model) to $1$ (Ising model). Moreover, we impose
periodic boundary conditions (PBC);
 $\sigma_{N}^{x}=\sigma_{0}^{x}$ and $\sigma_{N}^{y}=\sigma_{0}^{y}$. The zero-temperature phase
diagram of the model is schematically shown in
Fig.~(\ref{fig:phasediagram}). In free-field XY chain ($h=0$), the
system exhibits ferromagnetic order with non-zero magnetization
originating from the exchange coupling between nearest-neighbor
spins. Adding the external field tends to align the spins in the
$z$-direction such that the system undergoes a ferromagnet to
paramagnet transition at $h=1$. The green solid line $h=1$
represents the (Ising) critical points separating the regimes of
ferromagnetic and paramagnetic phase. The other critical green
solid line $r=0$ (isotropic model) is the boundary between the
ferromagnetic phases in $x$-phase (upper half-plane) and $y$-phase
(lower half-plane). The intersection of these two lines ($r=0$ and
$h=1$) is the XX critical point with a different universality
class from that of Ising universality. The ferromagnetic phase is
divided into two parts by the dashed blue circle: outside the
circle the correlation functions decay exponentially, while they
have oscillatory tails inside [\onlinecite{ref:BMD70}]. The ground
state on the dashed blue circle (called classical line) has a
simple direct product form of single qubit states implying zero
two-point functions and extremely short correlation length
[\onlinecite{ref:SS}].
\par
Quite generally, correlations are expected to reach a maximum as
long-range correlations dominate system's behavior at the critical
point. QPT occurs when small variation in the parameters of
Hamiltonian fundamentally changes the symmetry of the ground
state, resulting in the actual level-crossing or the limiting case
of avoided level crossing between the ground and excited state
[\onlinecite{ref:SQPT}]. At the critical point, the length scale
characterizing the exponential decay or the crossover of
correlation functions diverges as $\xi\sim|h-h_c|^{-\nu}$. On the
other hand, entanglement, as a purely quantum mechanical property,
has a close relation with quantum correlations and can be
exploited as an indicator of quantum phase transition
[\onlinecite{ref:TWM},\onlinecite{ref:OAFF02}]. However,
bi-partite measures of entanglement between individual qubits
typically decay fast as a function of distance even near the
critical point
[\onlinecite{ref:OAFF02},\onlinecite{ref:GTLC},\onlinecite{ref:GLL},\onlinecite{ref:VPM}].
It is therefore expected that a more global (multi-partite)
measure of entanglement would provide a more appropriate measure
to study criticality in QPT.

In this paper, we use global entanglement measure introduced by
Meyer and Wallach [\onlinecite{ref:MW}]. The measure is a function
of N-qubit pure states of $\ket{\psi}\in(\mathbb{C}^2)^{\otimes
N}$ as [\onlinecite{ref:brennen}]:
\begin{equation}\label{Egpsi}
  E_g(\ket{\psi})=\frac{4}{N}\sum_{k=1}^{N}D(\ket{\tilde{u}^{k}},\ket{\tilde{v}^{k}})
\end{equation}
where the non-normalized vectors $\ket{\tilde{u}^{k}}$ and
$\ket{\tilde{v}^{k}}$ are the projections of the state
$\ket{\psi}$ onto the \textit{$k^{th}$} qubit subspaces:
\begin{equation}\label{psi}
\ket{\psi}=\ket{0}\otimes\ket{\tilde{u}^{k}}+\ket{1}\otimes\ket{\tilde{v}^{k}}
\end{equation}
and $D$ is the norm-squared of the wedge product of two vectors $\ket{\tilde{u}^{k}}$ and $\ket{\tilde{v}^{k}}$ as
\begin{equation}\label{Ddefine}
D(\ket{\tilde{u}^{k}},\ket{\tilde{v}^{k}})=\sum_{i<j}\mid
\tilde{u}_{i}^{k}\tilde{v}_{j}^{k}-\tilde{u}_{j}^{k}\tilde{v}_{i}^{k}\mid^{2}.
\end{equation}
Meyer and Wallach [\onlinecite{ref:MW}] proved that this measure
is entanglement monotone in the sense that it is non-increasing
function under local operations and classical communications
(LOCC). Using the invariance under local operations, $\ket{\psi}$
can be written in the Schmidt basis over bipartite divisions of
the \textit{$k^{th}$} qubit and other qubits that simplifies Eq.
(\ref{Egpsi}) as [\onlinecite{ref:brennen}]:
\begin{equation}\label{GE}
  E_g=2\Big(1-\frac{1}{N}\sum_{k=0}^{N-1}tr(\rho_k^2)\Big)
\end{equation}
where $\rho_k$ is the reduced density matrix for the $k^{th}$
qubit obtained by tracing over other qubits. Global entanglement
($E_g$) has been used to detect quantum critical points
[\onlinecite{ref:oliveira}], and its scaling properties for the
Ising model in various dimensions has been studied
in [\onlinecite{ref:montakhab}]. It has also been used to study
decoherence effects in finite qubit systems
[\onlinecite{ref:montakhab2}].
\par

Since the MW measure was unable to distinguish global and
sub-global entanglements (e.g.~globally entangled 4-qubit state
and product of two 2-qubit entangled
states [\onlinecite{ref:LBS}]), Scott [\onlinecite{ref:Scott}]
generalized the MW measure to multi-qudit states of
$\ket{\psi}\in(\mathbb{C}^D)^{\otimes N}$ considering all the
possible bipartite divisions as
\begin{equation}\label{CGE}
 Q_{m}(\psi)=\frac{D^m}{D^{m}-1}\Big(1-\frac{m!(N-m)!}{N!}\sum_{\mid S\mid=m}tr(\rho_S^2)\Big)
\end{equation}
in which $m=1,2...,[N/2]$  ($[k]$ denotes the integer part of $k$)
and $S$ is a set of $m$-qubits. Although $Q_{m}$ is able to
distinguish between fully global and sub-global entanglements, its
direct computation is a quite challenging task for realistic
models as it requires all $m$-site reduced density matrices.  We
therefore propose to study the XY model using $E_g$ to see how
much information one can extract regarding various transitions
using such a measure.

\section{Ising Quantum Phase Transition}
\subsection{XY Model } \label{sec:XY}
\begin{figure}
  \center
   \includegraphics[width=0.4\textwidth]{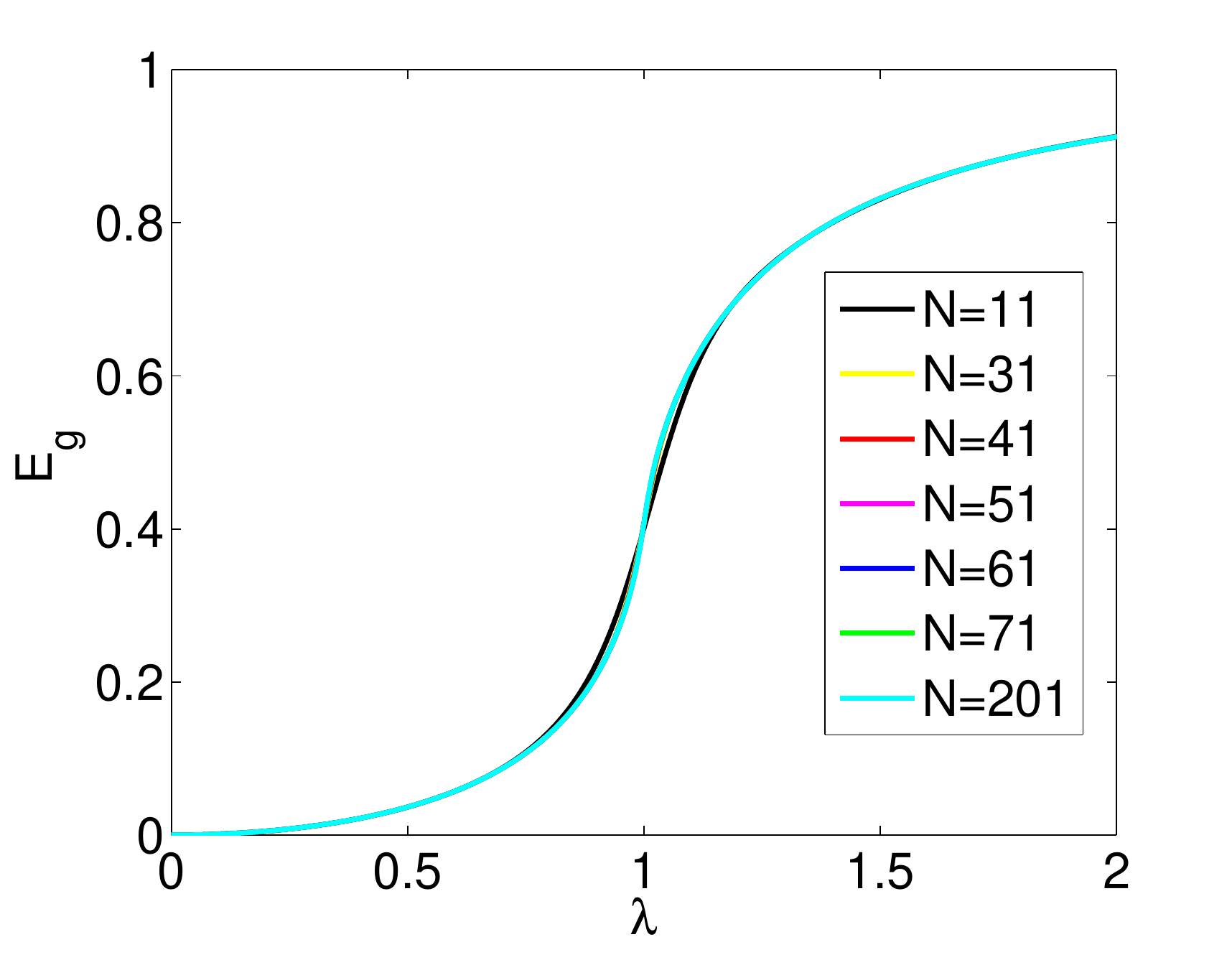}
  \caption{\label{fig:EgLambda}Global entanglement for a transverse XY chain ($r=0.5$) as a function of Hamiltonian parameter $\lambda=J/h$ for various system sizes.}
\end{figure}

\begin{figure}[h]
  \center
   \includegraphics[width=0.5\textwidth]{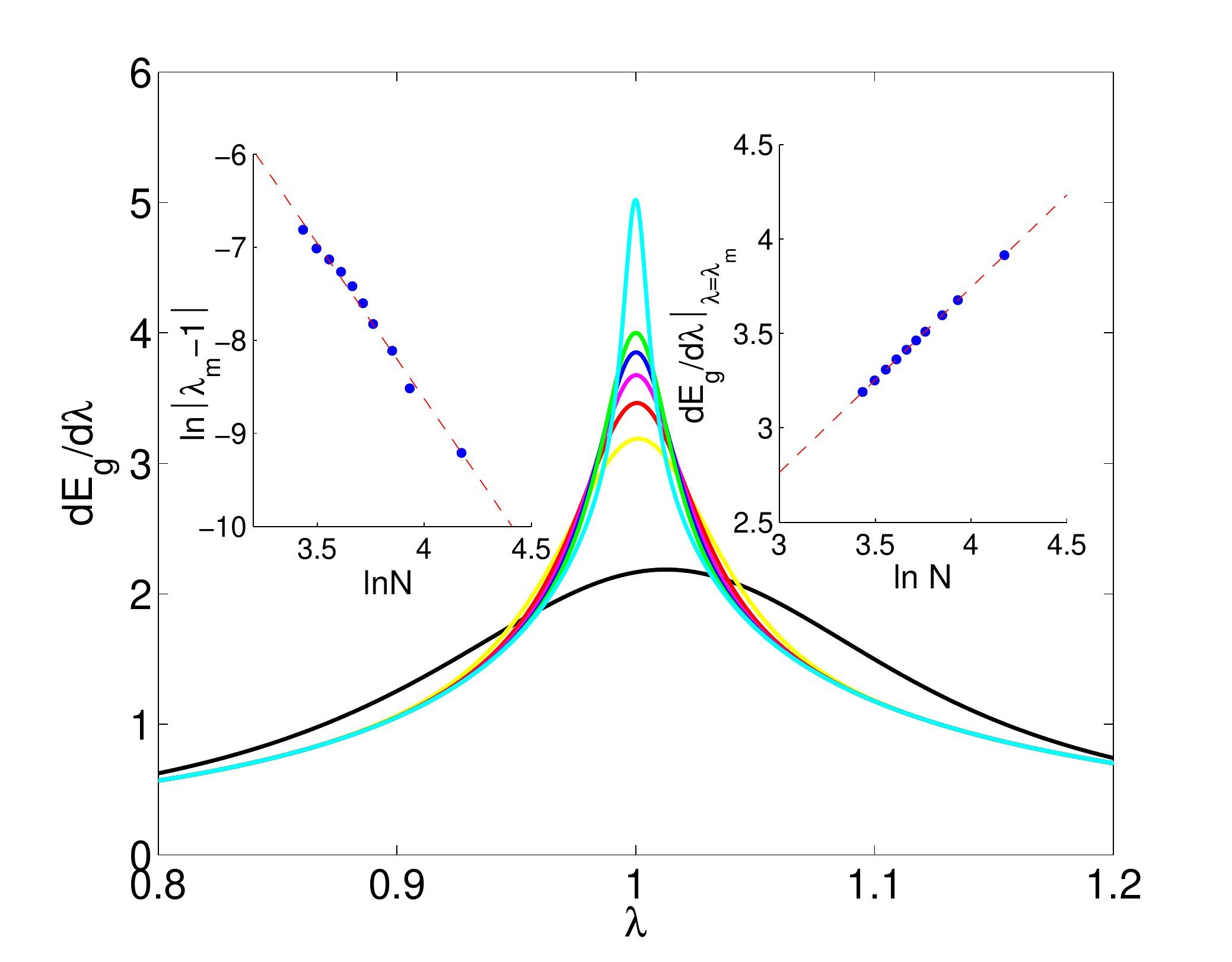}
  \caption{\label{fig:dEg}The derivative of global entanglement for a transverse XY chain ($r=0.5$) as a function of Hamiltonian parameter $\lambda$ for various system sizes.
  The left inset shows that the maximal value $\lambda_m$ approaching the critical point $\lambda_c=1$ as $|\lambda_m-\lambda_c|\sim N^{-3.34}.$
  The right inset shows the logarithmic divergence of the peak as a function of $N$, $\frac{dE_g}{d\lambda}\mid_{\lambda_m}\approx \kappa_1 \ln N$ where $\kappa_1=0.9783$.
   The system sizes are the same as Fig.~(\ref{fig:EgLambda}).}
\end{figure}

The transverse XY Heisenberg model has been solved in
[\onlinecite{ref:LSM}] by the Jordan-Wigner transformation which
maps the spin operators $\sigma_i$ into the spinless fermionic
operators:
\begin{align}\label{JWtransform}
        &\sigma_i^z=1-2c_i^{\dag}c_i&&\sigma_{i}^{\dag}=(\Pi_{j<i}\sigma_{j}^z)c_i,
       \end{align}
followed by Fourier transformation [\onlinecite{ref:Zhu}]
\begin{equation}\label{FT}
c_k=\frac{1}{\sqrt{N}}\sum_{j=0}^{N-1}\exp(-\frac{2\pi ikj}{N})c_j
\end{equation}
and Bogoliubov transformation
\begin{equation}\label{BT}
b_k=c_k\cos(\theta_k/2)-ic^{\dag}_{-k}\sin(\theta_k/2)
\end{equation}
where $\cos\theta_k=\frac{\cos(2\pi k/N)-h}{\omega_k}$ and
\begin{equation}\label{omegak}
\omega_k=\sqrt{\Big(h-\cos(2\pi k/N)\Big)^2+r^2\sin^2(2\pi k/N)}.
\end{equation}
Thus, the Hamiltonian takes the diagonal form
\begin{equation}\label{diagonalH}
H=\sum_{k}\omega_k(b_{k}^{\dag}b_k-1)
\end{equation}
where we may neglect the boundary terms for large systems. We want
to obtain an analytical expression for global entanglement of the
XY chain in presence of traverse magnetic field. To this end, we
expand the reduced density matrix $\rho_{i}$ as
[\onlinecite{ref:osborne}]:
 \begin{equation}\label{rhosingle1}
\rho_{i}=\frac{1}{2}\Big(q_{0}I+\sum_{\mu=x,y,z}q_{\mu}\sigma_{i}^{\mu}\Big).
\end{equation}
where $I$ is the identity matrix and $\sigma_{i}^{\mu}(\mu=x,y,z)$ are the Pauli matrices at $i^{th}$ qubit.
The reality of the Hamiltonian (Eq.\eqref{hamiltonian}) and its
global phase-flip symmetry ($[\Pi_{i=0}^{N-1}\sigma_{i}^{z},H]=0$)
implies $q_{x}=q_{y}=0$. In addition, the reduced density matrix
is unit-trace; so $q_{0}=1$ and the single particle density matrix
can be written as
[\onlinecite{ref:osborne},\onlinecite{ref:BM71}]:
\begin{equation}\label{rhosingle2}
\rho_{i}=\frac{1}{2}\Big(I+\langle\sigma_{i}^{z}\rangle\sigma_{i}^{z}\Big),
\end{equation}
where
\begin{align}\label{rhosingle3}
&\langle\sigma_{i}^{z}\rangle=-\frac{2}{N}\sum_{k=1}^{\frac{N-1}{2}}\frac{\cos(2\pi k/N)-h}{\sqrt{r^{2}\sin^{2}(\frac{2\pi k}{N})+[h-\cos(\frac{2\pi k}{N})]^{2}}}-\frac{1}{N}
\end{align}
and we consider odd number of qubits. Since we use PBC, the single
particle density matrix is the same for all the qubits of chain
and using Eq.\eqref{GE} we get [\onlinecite{ref:montakhab}]:
\begin{equation}\label{GE2}
  E_g = 2(1-tr\rho_i^2)
\end{equation}
and the global entanglement in this case can be obtained as
\begin{equation}\label{Eglf}
E_{g}=1-\langle\sigma_{i}^{z}\rangle^{2}
\end{equation}
We begin our analysis by considering $E_g$ as a function of
$\lambda=\frac{J}{h}$ for various system sizes $N$, and fixed
value of $r=0.5$. The results are shown in Fig.~(\ref{fig:EgLambda}).
$E_g$ increases from zero at $\lambda=0$ ($h\rightarrow \infty$),
where the ground state of the system is a product state of spins
aligned in the $z$-direction and reaches  its maximal value
$E_g=1$ with a sharp rise at (the finite-size) transition point,
$\lambda_m$. A better picture arises when one looks at the
derivative of such function which displays a divergence at the
critical point in the thermodynamic limit. Fig.~(\ref{fig:dEg})
displays such information. As the system size grows the peak of
the derivative approaches the critical point as it diverges in its
value.  As indicated in the inset the divergence is logarithmic
indicating a slow divergence. However, the convergence to the
critical point is fast as $\lambda_m$ approaches the critical
point $\lambda_c=1$ with $|\lambda_m-\lambda_c|\sim N^{-\alpha}$
with relatively large ``finite-size exponent" of $\alpha=3.34$. As
indicated in the inset, the maximum value of the derivative
$dE_g/d\lambda$ obeys:
\begin{equation}\label{dEglambdam}
\frac{dE_g}{d\lambda}\mid_{\lambda_m}\approx \kappa_1 \ln N.
\end{equation}
with $\kappa_1=0.9783.$

\begin{figure}[h]
  \center
   \includegraphics[width=0.4\textwidth]{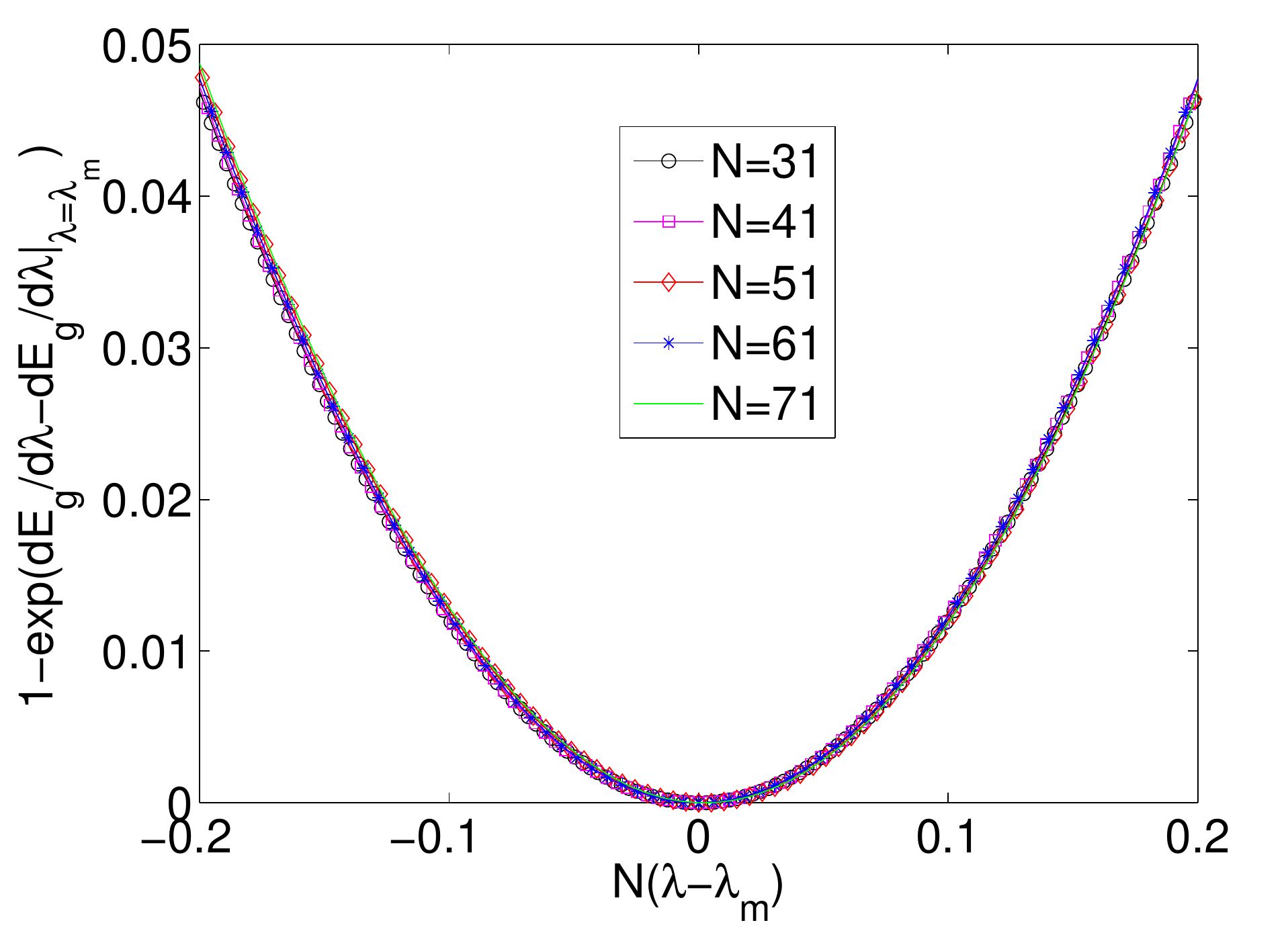}
  \caption{\label{fig:expGElambda}Finite-size scaling data collapse of derivative of global entanglement for transverse XY chain ($r=0.5$). The best collapse of $\frac{dE_g}{d\lambda}\sim Q(N^{1/\nu}(\lambda-\lambda_m))$ occurs at $\nu=1$. }
\end{figure}

We next calculate the all-important critical exponent $\nu$ using
finite size scaling of the derivative of $E_g$. The scaling
relation we use is $\frac{dE_g}{d\lambda}\sim
Q(N^{1/\nu}(\lambda-\lambda_m))$ with $Q(x)\sim ln(x)$
[\onlinecite{ref:MNB}]. As can be seen in
Fig.~(\ref{fig:expGElambda}), all the curves of
$F=1-\exp[\frac{dE_g}{d\lambda}-\frac{dE_g}{d\lambda}\mid_{\lambda=\lambda_m}]$
as a function of $N^{1/\nu}(\lambda-\lambda_m)$ collapse nicely on
a single curve for $\nu=1$, in agreement with the well known
result for the Ising universality class
[\onlinecite{ref:LSM},\onlinecite{ref:McCoy}]. We finally note
that our results was obtained for $r=0.5$, however, similar
results hold for $0<r\leq1$

\subsection{XX Model}
XX model is the isotropic case of XY Heisenberg model which
belongs to a different universality class from that of the Ising.
In this model, as the magnetic field is varied, the energy gap
between the ground and the first excited state vanishes and the
intersections exhibit a sequence of level-crossing points for the
finite-size chains . Since the global entanglement directly
depends on the ground state of the system, we expect to see sudden
jumps in $E_g$ at the level-crossing points. To this end, we are
interested in the global entanglement behavior for finite-size
chains where the boundary effect terms of the  Hamiltonian cannot
be neglected. These terms break the periodicity of the
Jordan-Wigner operators
\begin{equation}\label{cpJW}
c_{i}^{\dag}=\big(\Pi_{j<i}\sigma_{j}^{z}\big)\sigma_{i}^{\dag}=e^{i\pi n_{i}\downarrow}\sigma_{i}^{\dag}
\end{equation}
as
\begin{align}\label{cpJWPB}
        &c_{0}^{\dag}=\sigma_{0}^{\dag}&&c_{N}^{\dag}=e^{i\pi n_{N}\downarrow}\sigma_{N}^{\dag}=e^{i\pi n_{N}\downarrow}\sigma_{0}^{\dag},
       \end{align}
in which $n_{i\downarrow}$ is the operator counting the total
number of spin-down in the chain. In this case, the Hamiltonian
can be diagonalized by the Jordan-Wigner transformation and the
following deformed Fourier transformation
[\onlinecite{ref:PCF},\onlinecite{ref:PF}]:
\begin{equation}\label{DFT}
c_j=\frac{1}{\sqrt{N}}e^{\frac{2\pi i \alpha_j}{N}}\sum_{k}e^{-ikj}c_k
\end{equation}
where $\alpha_j$ is a local gauge. The ground state of this model was obtained in [\onlinecite{ref:PCF}] as
\begin{eqnarray}\label{DFT}
\ket{\psi_{n}}&=&\frac{1}{\sqrt{N}}\sum_{j_1<j_2<...<j_n}\left\{\lambda_{j_1,j_2,...,j_n}(-1)^{nj_1}(-1)^{(n-1)(j_2-j_1)}\right.\cr\cr
&&\left.(-1)^{(n-2)(j_3-j_2)}
...(-1)^{j_n-j_{n-1}}\right\}\ket{\downarrow}_{0}...\ket{\uparrow}_{j_1}...\ket{\uparrow}_{j_2}\cr\cr
&& ...\ket{\uparrow}_{j_n}...\ket{\downarrow}_{N-1}.
\end{eqnarray}
while $\lambda_{j_1,j_2,...,j_n}$ is given by
\begin{equation}\label{lambdaxx}
\lambda_{j_1,j_2,...,j_n}=\sum_{p}(-1)^{p}\exp\Big(\frac{2\pi i}{N}(k_1j_{p_1}+k_2j_{p_2}+...+k_nj_{p_n})\Big)
\end{equation}
and $1\leq n\leq [N/2]$ depends on the magnetic field $h$ such that
\begin{eqnarray}\label{EgXXh}
\frac{\sin[\frac{(n+1)\pi}{N}]-\sin(\frac{n\pi}{N})}{\sin(\pi/N)}<
h\leq\frac{\sin(\frac{n\pi}{N})-\sin[\frac{(n-1)\pi}{N}]}{\sin(\pi/N).}
\end{eqnarray}
Moreover, the sum (in Eq. \eqref{lambdaxx}) extends over the permutation group. Therefore, the $z$
magnetization is $\langle\sigma_{i}^{z}\rangle=1-\frac{2n}{N}$.
This allows us to calculate global entanglement in an analytic
fashion for finite size system, which leads to:
\begin{eqnarray}\label{EgXX}
E_{g}=\frac{4n(N-n)}{N^2}
\end{eqnarray}

\begin{figure}[h]
\center
\includegraphics[width=0.4\textwidth]{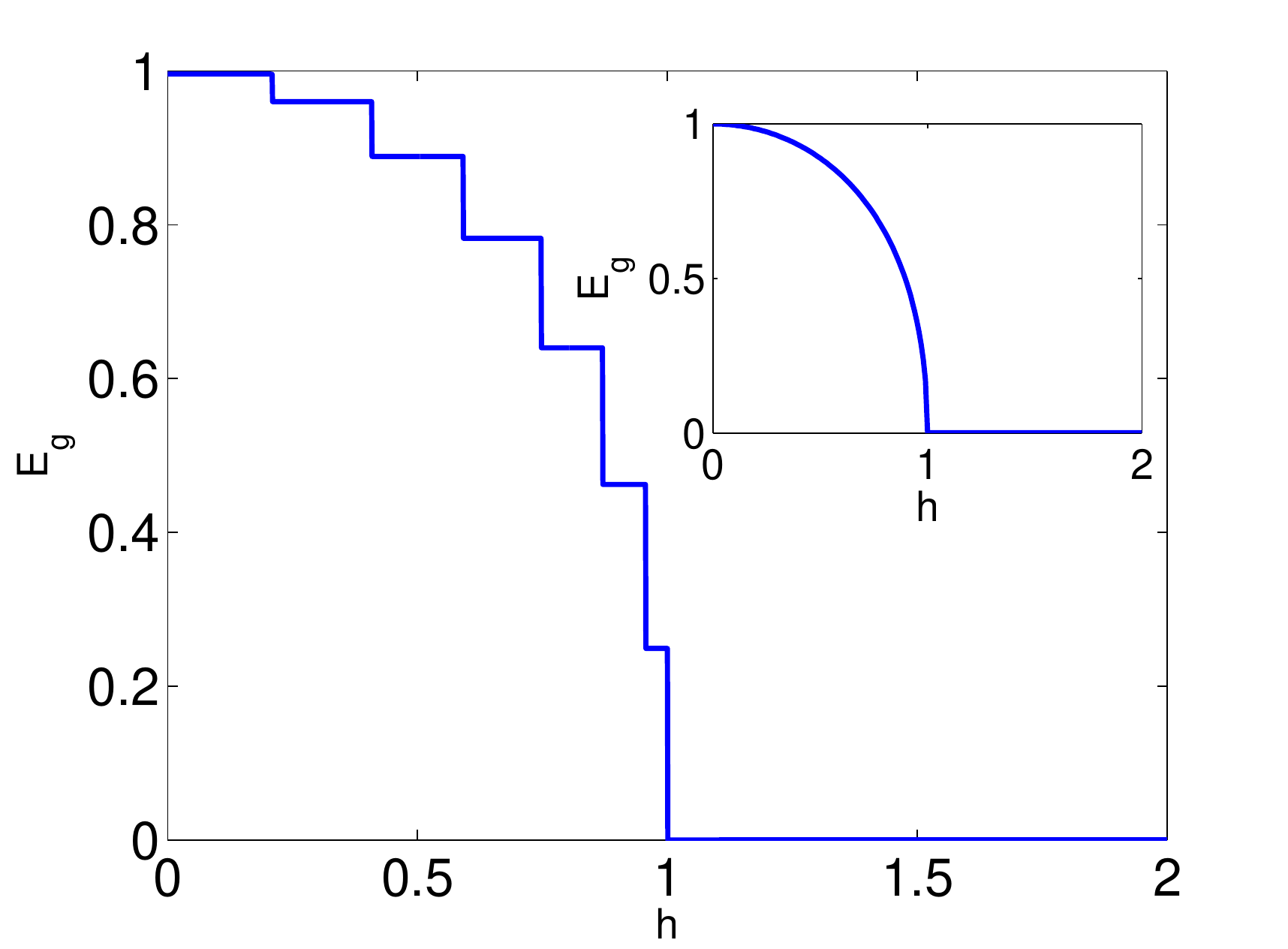}
\caption{\label{fig:XXd1}Global entanglement (Eq. (\ref{EgXX})) as
a function of magnetic field for XX Heisenberg ($r=0$) chain of
$N=15$ qubits. The level-crossings coincide with jumps in $E_g$.
The inset shows the $N \rightarrow \infty$ limit.}
\end{figure}

Interestingly, this indicates a step-like behavior for the $E_g$
as a function of $h$. In fact, the points
$h_{s}=\frac{\sin(\frac{n\pi}{N})-\sin[\frac{(n-1)\pi}{N}]}{\sin(\pi/N)}$
are exactly the same as the level-crossing points obtained by the
exact solution of XX model, see
[\onlinecite{ref:PCF},\onlinecite{ref:PF}]. Fig.~(\ref{fig:XXd1})
shows the global entanglement for a XX chain of $N=15$ qubits
obtained from  Eq.\eqref{EgXX} in terms of $h$. The stepwise
behavior of global entanglement determines the position and number
of level-crossings in the system. Note that the number of steps is
$[N/2]$.

\par
It might be interesting to look into the behavior of $E_g$ in the
thermodynamic limit for the XX model to see what happens to the
step like structure, as well as the behavior near the critical
point. In the thermodynamics limit, we can neglect the boundary
effects and use Eq.\eqref{rhosingle3} at $r=0$ and write the
global entanglement for XX model
\begin{eqnarray}\label{detailsEgXXTL}
E_{g}=1-\left\{\frac{1}{\pi}\int_{0}^{\pi}\frac{h-\cos(\phi)}{\mid h-\cos(\phi)\mid}\right\}^2=1-\frac{1}{\pi^2}\Big(2\phi_c-\pi\Big)^2
\end{eqnarray}
 where $\phi_c=\cos^{-1}(h)$ is the pole of the denominator. The
 behavior is shown as an inset in Fig.~(\ref{fig:XXd1}). The effect of finite number of
 steps naturally disappear and $E_g$
 behaves much as an order parameter for this transition.  One can
 also use this expression to obtain scaling of the derivative of
 entanglement in order to obtain correlation length exponent [\onlinecite{ref:TWM},\onlinecite{ref:Zhu}].
Hence, we get
\begin{eqnarray}\label{dEgXXTL}
\frac{dE_{g}}{dh}|_{h\rightarrow1^{-}}\approx
\frac{4\sqrt{2}}{\pi^2}\frac{1}{\sqrt{1-h}.}
\end{eqnarray}
This allows us to obtain the exponent $\nu$, which governs the
divergence of the correlation length as $\frac{dE_g}{dh}\sim|h-1|^{-\nu}$ with $\nu=1/2$, consistent with
previous reports [\onlinecite{ref:MNB},\onlinecite{ref:McCoy}].

\section{Thermodynamic limit and the classical line}

\begin{figure}[h]
  \center
   \includegraphics[width=0.4\textwidth]{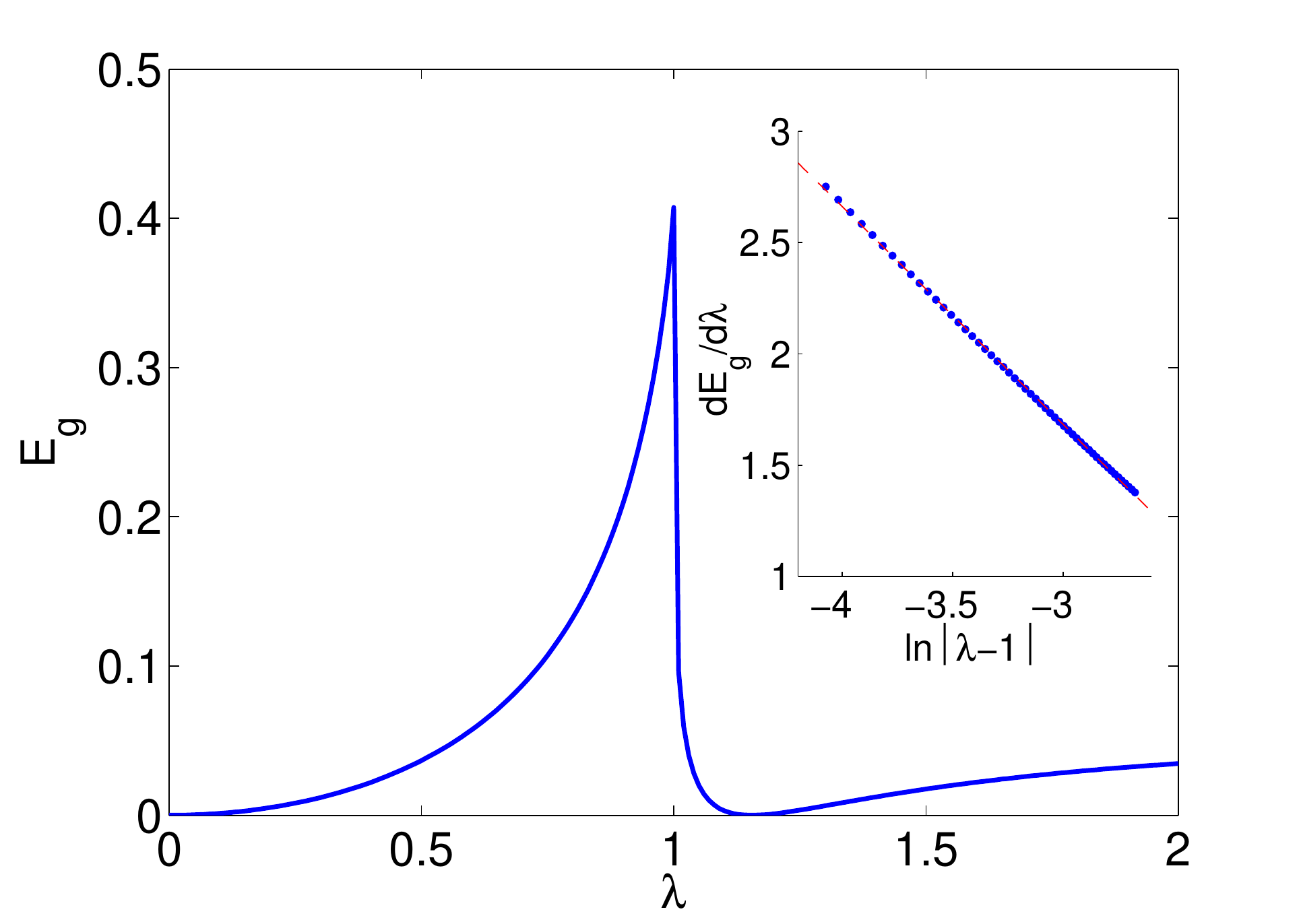}
  \caption{\label{fig:GElambdaTL}Global entanglement as a function of $\lambda$ for $r=0.5$ in the thermodynamic limit.
  The inset displays the logarithmic divergence behavior of $dE_g/d\lambda$ as it approaches the critical point. The slope gives $\kappa_2=-0.9789$.}
\end{figure}

In the previous section, we used the finite-size behavior of $E_g$
in order the characterize the behavior of QPT at $h=1$, as well as
characterizing the XX model.  We also obtained a closed form
expression for $E_g$ in the thermodynamics limit which allowed us
to calculate the critical exponent for this universality class. We
now propose to calculate $E_g$ in the thermodynamic limit for the
entire parameter regime and extract more information in this limit
of the system, paying particular attention to the so-called
classical transition.  Let us start by explaining the simplest
model of XY Heisenberg family, the Ising model ($r=1$). In the
absence of external field ($h=0$), the spins are either all
pointed in the positive or negative $x$ direction and the
corresponding ground state is doubly degenerate. Turning on a
small $h$ may be regarded as a perturbation changing the
orientation of a small fraction of spins to the opposite
direction. In the case of finite size system, such a field induces
quantum tunneling between the degenerate ordered states and leaves
the system in a superposition state satisfying phase-flip
symmetry. As we go to the thermodynamic limit, this energy barrier
becomes infinitely high such that it does not allow tunneling
events for any finite $h$ and therefore keeps the system in the
degenerate ground state [\onlinecite{ref:JLEI}]. Therefore, this
breaking of phase-flip symmetry requires us to take into account
the coefficient $q_x$ in Eq.\eqref{rhosingle1} and rewrite
Eq.\eqref{rhosingle2} as
[\onlinecite{ref:osborne},\onlinecite{ref:Pfuty}]:
\begin{align}\label{rhosingleTL}
\rho_{i}=\frac{1}{2}\Big(I+\langle\sigma_{i}^{z}\rangle\sigma_{i}^{z}+\langle\sigma_{i}^{x}\rangle\sigma_{i}^{x}\Big)
\end{align}
where
\begin{equation}\label{sigmazint}
\langle \sigma^{z}\rangle=\frac{1}{\pi}\int_{0}^{\pi}d\phi\frac{h-\cos(\phi)}{\sqrt{r^2\sin^{2}(\phi)+[h-\cos(\phi)]^2}},
\end{equation}
and
\begin{equation}\label{sigmaxint}
\langle\sigma^x\rangle=
  \begin{cases}
    2[2(1+r)]^{-1/2}r^{1/4}(1-h^2)^{1/8}     & \quad \text{if } h\leq1\\
    0   & \quad \text{otherwise}
  \end{cases}
\end{equation}
Given the above and considering Eq.\eqref{GE2}, global
entanglement can now be obtained for the entire parameter regime
in the thermodynamic limit. Fig.~(\ref{fig:GElambdaTL}) shows an
example for $E_g$ as a function of $\lambda$ for $r=0.5$.

\begin{figure}[h]
  \center
   \includegraphics[width=0.4\textwidth]{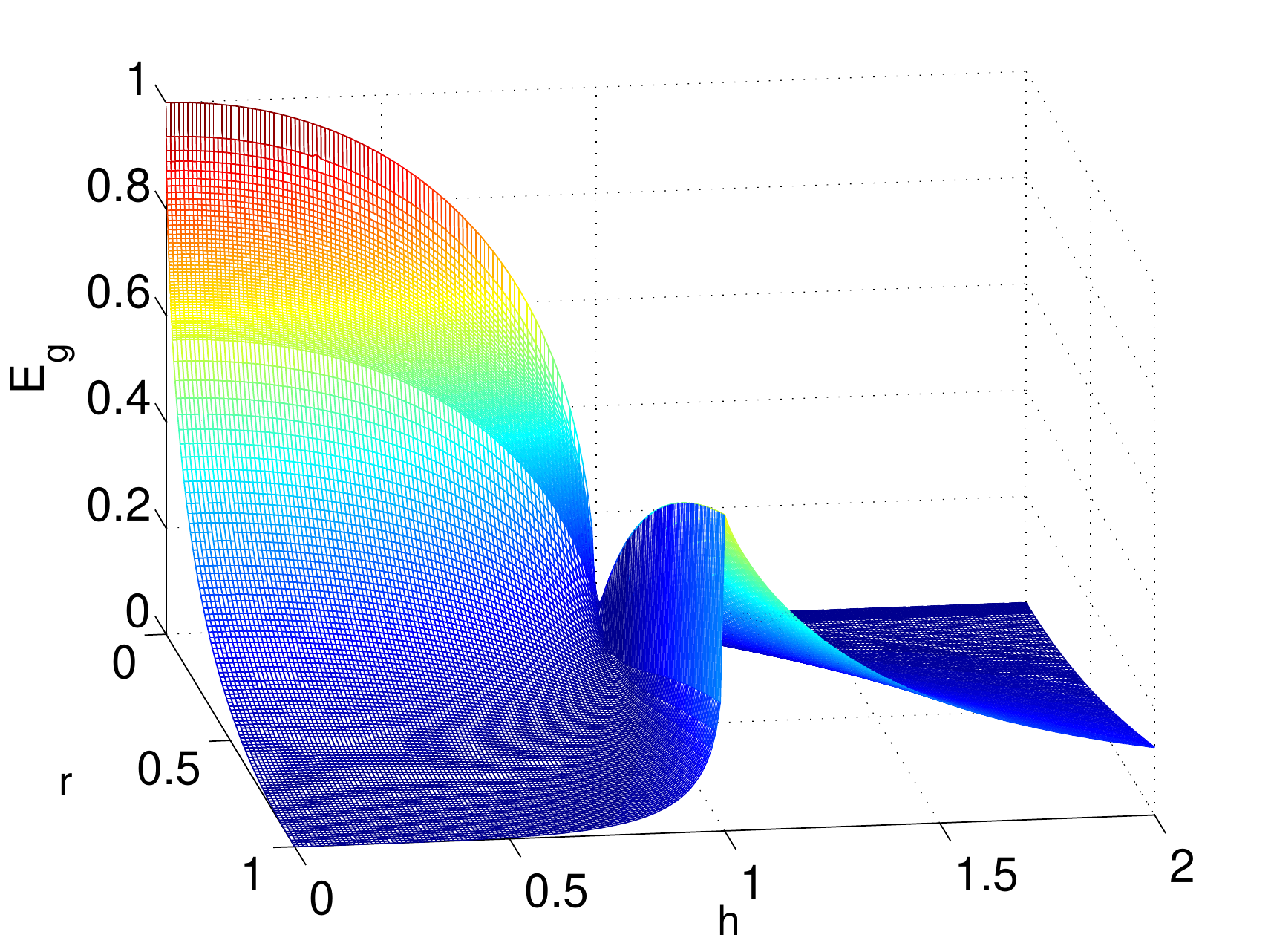}
  \caption{\label{fig:GErhTL}3D plot of global entanglement as a function of $r$ and $h$ in the thermodynamic limit.}
\end{figure}

\begin{figure}[h]
  \center
   \includegraphics[width=0.4\textwidth]{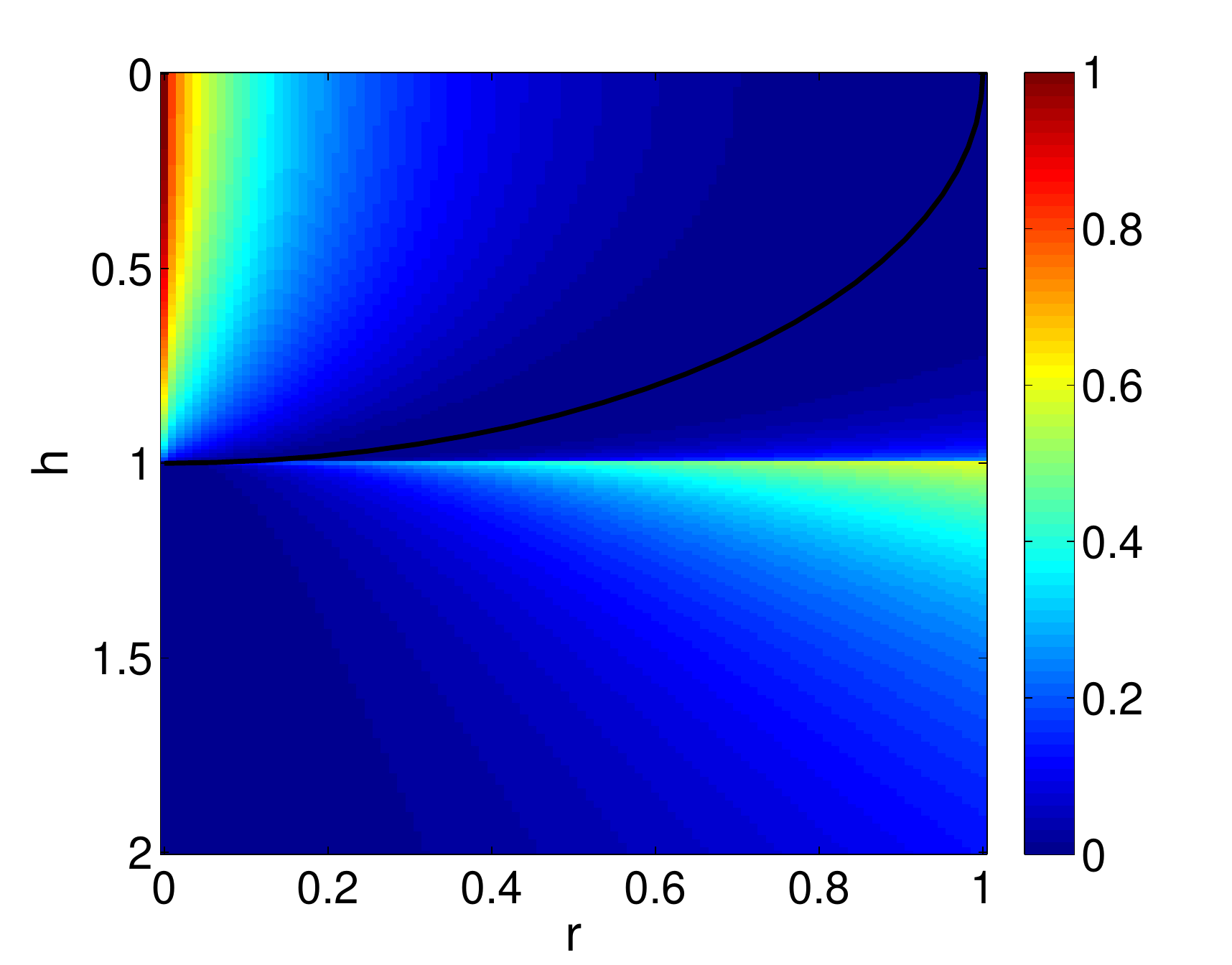}
  \caption{\label{fig:CPGErhTL}Contour plot of global entanglement versus the external field and anisotropic parameter.}
\end{figure}

In the weak exchange regime ($\lambda<1$), the XY Hamiltonian term
may be regarded as a perturbation that is unable to break the
phase-flip symmetry and therefore leaves the system in
non-degenerate ground state, so $E_g$ in this case is the same as
the one for finite-size chains (see Fig.~(\ref{fig:EgLambda})). At
the critical point, $\langle\sigma^x\rangle$ begins to rise,
breaking the phase-flip symmetry, leading to a sudden decline in
entanglement and exhibiting absolute maximal value for
entanglement at the critical point. This by the way is consistent
with the general expectation of highly correlated system at the
critical point [\onlinecite{ref:oliveira},\onlinecite{ref:osborne}]. However, entanglement quickly
decreases and vanishes at $\lambda=1.15$ ($h=0.87$) where the
ground state is unentangled (product states) since it lies exactly
on the classical line $r^2+h^2=1$.  In order to obtain a better
understanding of the behavior of $E_g$, we plot it in terms of $h$
and $r$ in Fig.~(\ref{fig:GErhTL}).
For all the non-zero anisotropic parameter, $E_g$ is maximum on
the Ising transition line separating paramagnetic and
ferromagnetic phases. In the case of $r=0$ (isotropic model), the
global entanglement exhibits a different behavior indicating a
different universality class. For a better understanding, a
corresponding contour plot is provided in Fig.~(\ref{fig:CPGErhTL}).
Here, we can more clearly see the behavior on the classical line.
The solid black line separates the oscillation part of
ferromagnetic phase in the diagram phase where the ground state is
a product state and $E_g$ is exactly zero along this transition.
Therefore, $E_g$ is also able to indicate the transition to the
oscillating phase in the ferromagnetic case as it becomes zero
across such transition.  We note that this value of zero, and
therefore indicator of the classical transition, is only valid in
the thermodynamic limit and does not occur for the finite size
systems. Note also that in the thermodynamic limit entanglement is
maximal at the critical point as expected, but only exhibits a
sharp (well-behaved) rise for the finite $N$ even if $N$ is taken
to be very large.  This provides a good indication that $E_g$ can
behave similar to the usual thermodynamic functions, as they
\emph{only} exhibit \emph{non-analytic} behavior (at criticality)
in the thermodynamic limit, compare Fig.~(\ref{fig:EgLambda}) with
Fig.~(\ref{fig:GElambdaTL}).  Moreover, the inset in
Fig.~(\ref{fig:GElambdaTL}) shows that the derivative of global
entanglement $dE_g/d\lambda$ for an infinite chain diverges as:
\begin{equation}\label{dEglambdathermo}
\frac{dE_g}{d\lambda}\approx\kappa_2 \ln \mid\lambda-1\mid.
\end{equation}
where $\kappa_2=-0.9789$. Based on the scaling ansatz for
logarithmic divergence [\onlinecite{ref:MNB}], the ratio of
$\mid\kappa_1/\kappa_2\mid$ is the correlation length exponent,
$\nu$. In our case, this ratio is given by
$\mid\kappa_1/\kappa_2\mid=0.994$ which is very close to the exact
result $\nu=1$ as well as our result in Sec. IIIA. We also note
similar results are obtained using concurrence in
[\onlinecite{ref:osborne}] and geometric phase in
[\onlinecite{ref:Zhu}].

\section{Anisotropic quantum phase transition}
\begin{figure}[h]
  \center
   \includegraphics[width=0.5\textwidth]{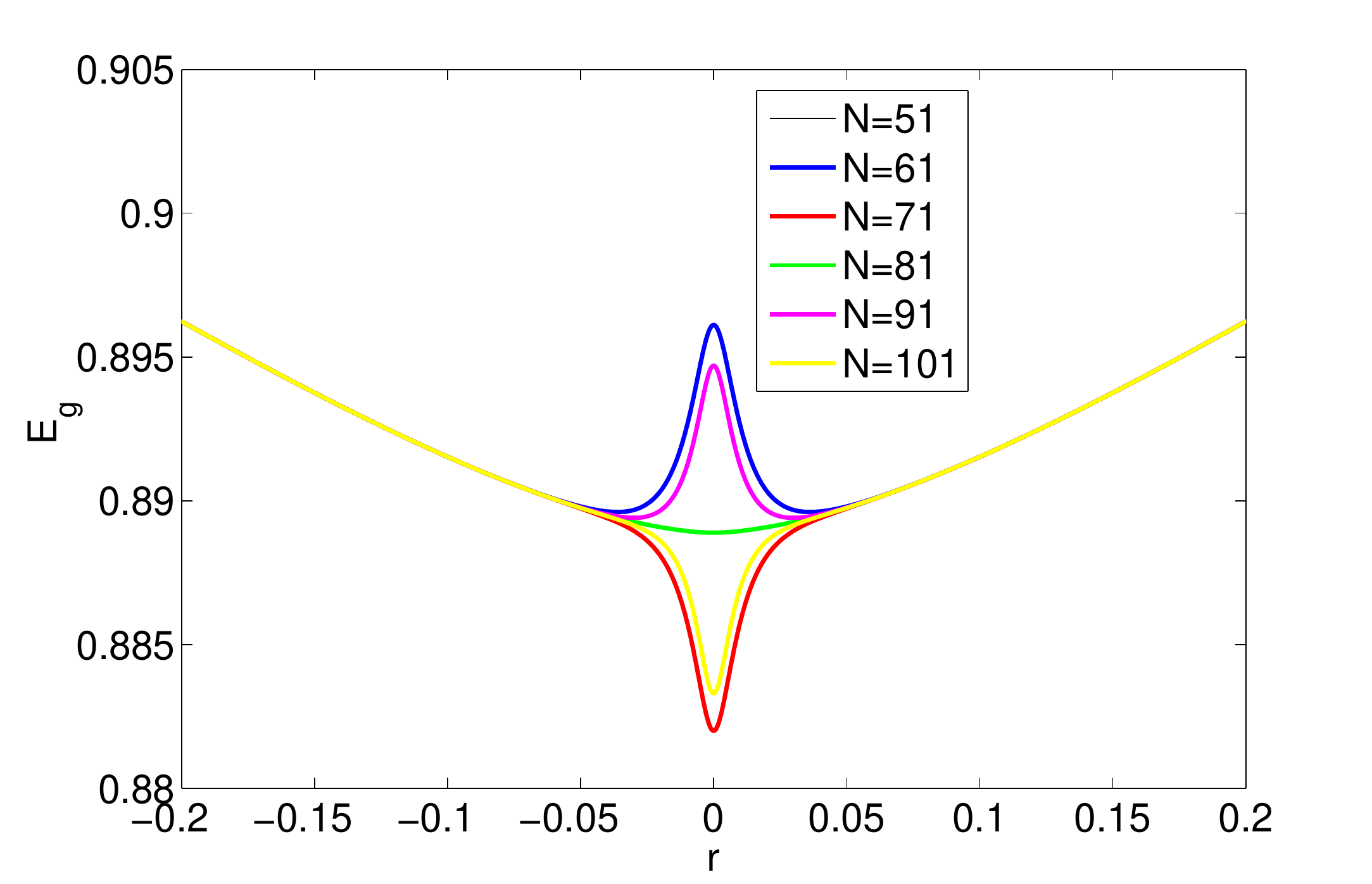}
  \caption{\label{fig:EgATFSNs}Global entanglement versus $r$ for the fixed magnetic field ($h=0.5$) and different system sizes.
  As $N$ changes, the step structure of the finite $XX$ model changes leading to the behavior observed.}
\end{figure}

Another quantum phase transition occurs over the line $r=0$ for
$0<h<1$, the anisotropic transition, which has received less
attention in the literature
[\onlinecite{ref:MRBD},\onlinecite{ref:CL},\onlinecite{ref:MZPT}].
The anisotropic phase transition separates two ferromagnetic
phases with orderings in the $x$ ($r>0$) and $y$ ($r<0$)
directions and belongs to a different universality class than the
Ising class. Fig.~(\ref{fig:EgATFSNs}) shows global entanglement as
one crosses such a transition for the fixed value of $h=0.5$.
As can be seen in this figure, there is a distinct change in the
global entanglement around $r=0$, which is the local extremum.
However, it may be minimum or maximum, depending on the system
size. We have observed that when  one is close but below (above)
the level crossing point, $E_g$ is convex (concave), slowly
changing shape as one crosses a given step for a fixed $N$.
Therefore, the picture that emerges is that for a finite chain,
the first derivative of $E_g$ is zero at the anisotropic
transition indicating a local maximum or a minimum depending on
whether the given values of $h$ and $N$ places us near the left or
right edge of the step. This pattern continues to hold across the
anisotropic transition until one gets to the multi-critical point
($h=1, r=0$), where $E_g$ displays a local minimum approaching
zero in the thermodynamic limit, see for example
Fig.~(\ref{fig:XXd1}). This type of behavior continues to hold for
$h>1$. Note that $E_g$ does not approach zero with increasing $N$
as one crosses the anisotropic transition. Therefore, one can
conclude that the finite size behavior of $E_g$ distinguishes the
critical anisotropic transition. However, one would like to know
if $E_g$ exhibits any non-analytic behavior associated with
(critical) quantum phase transitions.  To investigate this we need
to calculate global entanglement across the anisotropic transition
in the thermodynamics limit.

\begin{figure}[h]
  \center
   \includegraphics[width=0.4\textwidth]{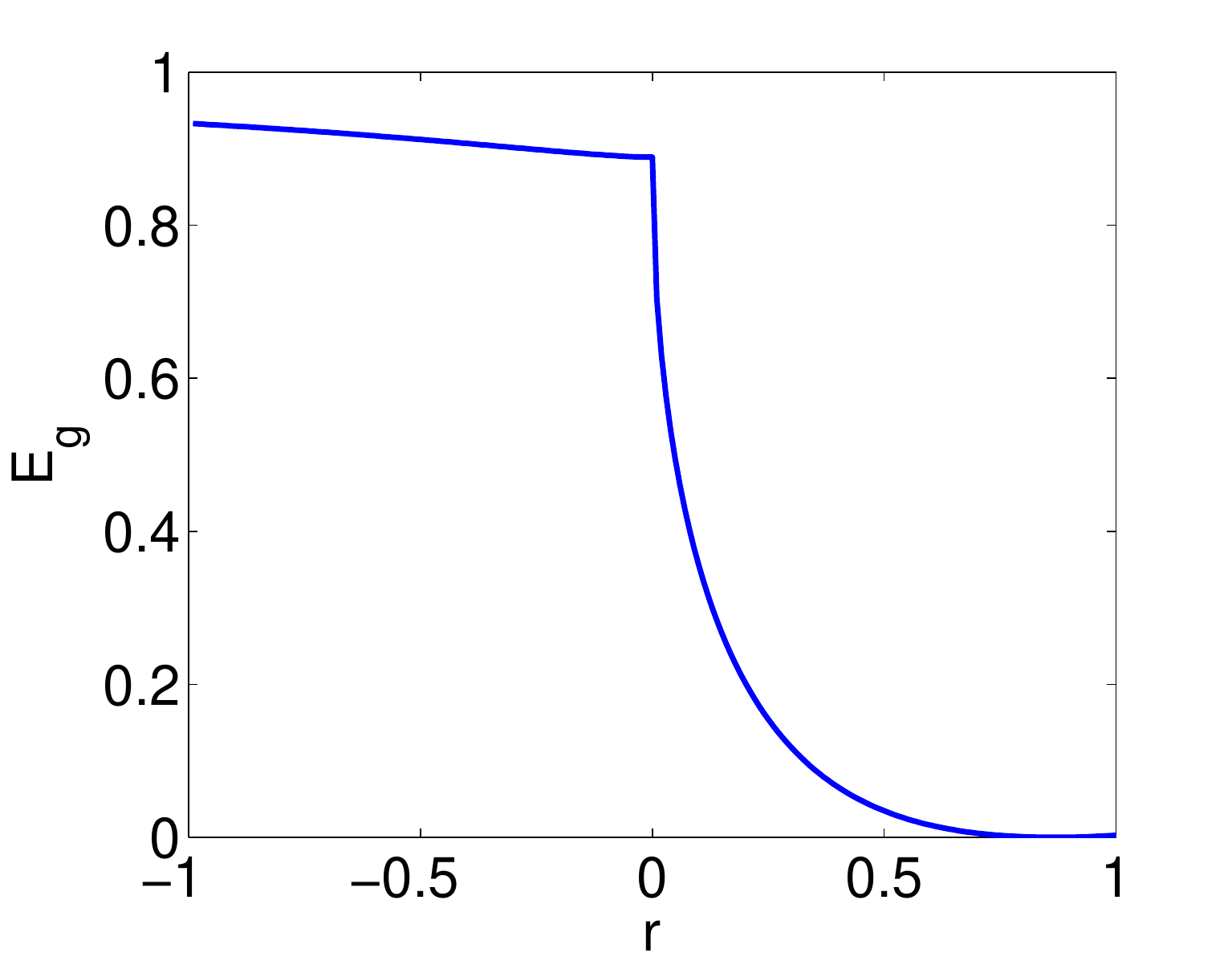}
  \caption{\label{fig:Egrh5} Global entanglement in terms of $r$ for a fixed value of $h=0.5$ in thermodynamic limit.
  }
\end{figure}

In order to calculate global entanglement for $r<0$ in the
thermodynamic limit we need to make the following observations. In
the regime of positive anisotropic parameter, magnetization in $y$
direction is zero and  $\rho_i$ is a function of $\sigma_x$ and
$\sigma_z$ as shown in Sec. IV, i.e. Eqs.\eqref{rhosingleTL},
\eqref{sigmazint}, \eqref{sigmaxint}. Therefore, $E_g$ is given by
Eq.\eqref{GE2}.  It is evident from the Hamiltonian that
transformation $r\rightarrow-r$ interchanges $\sigma_x$ with
$\sigma_y$, which leads to the zero value of
$\langle\sigma_x\rangle$ for $r<0$ [\onlinecite{ref:LSM}]. Therefore, since the
$y$-component of magnetization does not contribute to $E_g$, due
to the reality of Hamiltonian, single-particle density matrix is
given by $\rho_i=\frac{1}{2}(I+\langle\sigma_z\rangle\sigma_z$),
and $E_g$ is given by Eq.\eqref{Eglf} in this regime.  Therefore,
the picture that emerges is that for $h>1$ where one is in the
paramagnetic phase and no phase transition occurs at $r=0$, $E_g$
is symmetric about this minimum point for a given value of $h$.
However, one expects that the broken symmetry due to
$\langle\sigma_x\rangle$ at $r=0$ and $h<1$ leads to a broken
symmetry of $E_g$ about the transition point. This is indeed the
case as can be seen from Fig.~(\ref{fig:Egrh5}) which show $E_g$ for
$h=0.5$ across the anisotropic transition.  Clearly one can see
the non-analytic behavior similar to thermodynamic quantities at a
critical point. We conclude that $E_g$ can distinguish the
critical anisotropic transition in the $XY$ model.

\section{Conclusions}
In this paper, we have used global entanglement in order to study
various quantum phase transitions occurring in the transverse XY
Heisenberg chain.  We have been able to calculate global
entanglement both in the finite size limit as well as the
thermodynamic limit analytically.  The finite size study was shown
to be very useful for extracting the critical exponents for the
Ising transition, via the derivative of entanglement. In the
thermodynamic limit, $E_g$ was shown to exhibit non-analytic
behavior at the Ising transition, while having maximal value. The
thermodynamic limit of global entanglement was also used to
extract critical exponent for the multi-critical point of the XX
model. For the finite size system, the step structure of
level-crossings was exactly reproduced by global entanglement.
Also, while the finite size measure of global entanglement did not
show any particular behavior across the classical oscillating
transition, the thermodynamic limit of the measure was able to
signal such a transition as it took on vanishing value across this
classical transition. Furthermore, we studied the anisotropic
transition where global entanglement was shown to exhibit a
non-analytic behavior across such a transition in the
thermodynamic limit, while showing an interesting, $N$ dependent
behavior for the finite size case. The cusp-singularity at both
quantum transitions, non-analyticity at the multi-critical point
and vanishing value on the classical curve is the general behavior
of global entanglement in the thermodynamic limit. We therefore
conclude that global entanglement, despite its simplicity, can
produce much of the rich behavior of the XY model in various
parameter regimes, identifying all the transition points. Such a
multi-partite measure of entanglement seems to be a good candidate
for studying the thermodynamic behavior of many-body quantum
systems.

We end by making the following observation. We have seen that
while finite size study of entanglement can produce interesting
behavior including scaling properties, it was the thermodynamic
limit of entanglement which was able to fully bring to light the
various transitions in the XY model. In particular, the
thermodynamic limit exhibit properties which one would never see
even for very large values of $N$.  This is particularly relevant
as many studies of entanglement and quantum phase transitions are
limited by finite size studies with the belief that numerically
exact finite size solutions can be extrapolated to find the
thermodynamic limit.


\begin{references}

\bibitem{ref:SQPT} S. Sachdev, \textit{Quantum Phase Transitions},
Cambridge Univ. Press, Cambridge (2013).

\bibitem{ref:SGCS} S. L. Sondhi, S. M. Girvin, J. P. Carini, and D. Shahar, Rev. Mod. Phys. \textbf{69}, 315 (1997).

\bibitem{ref:HLG} D. B. Haviland, Y. Liu, and A. M. Goldman
Phys. Rev. Lett. \textbf{62}, 2180 (1989).

\bibitem{ref:PC} P. Coleman, Physica B, \textbf{259}, 353 (1999).

\bibitem{ref:NC} M. A. Nielsen and I. L. Chaung, \textit{Quantum Information and Quantum Computation}, Cambridge University Press, Cambridge (2000).

\bibitem{ref:OAFF02} A. Osterloh, L. Amico, G. Falci, and R. Fazio, Nature \textbf{416} 608-610 (2005).

\bibitem{ref:TWM} T. C. Wei, D. Das, S. Mukhopadyay, S. Vishveshwara, and P. M. Goldbart,
Phys. Rev. A \textbf{71}, 060305 (2005).

\bibitem{ref:CWHW} S. Chen, L. Wang, Y. Hao, and Y. Wang, Phys. Rev. A \textbf{77}, 032111 (2008).

\bibitem{ref:Zhu} S. L. Zhu, Phys. Rev. Lett. \textbf{96}, 077206 (2006).

\bibitem{ref:RD} R. Dillenschneider, Phys. Rev. B \textbf{78}, 224413 (2008).

\bibitem{ref:AK} T. Abad and V. Karimipour,
Phys. Rev. B \textbf{93}, 195127 (2016).

\bibitem{ref:GTLC} S. Gu, G. Tian, and H. Lin, Chin. Phys. Lett. \textbf{24} 2737 (2007).

\bibitem{ref:GLL} S. Gu, H. Lin, and Y. Li,
Phys. Rev. A \textbf{68}, 042330 (2003).

\bibitem{ref:VPM} J. Vidal, G. Palacios, and R. Mosseri,
Phys. Rev. A \textbf{69}, 022107 (2004).

\bibitem{ref:montakhab}
A. Montakhab, A. Asadian, Phys. Rev. A \textbf{82}, 062313 (2010).

\bibitem{ref:MW} D. A. Meyer and N. R. Wallach, Journal of Mathematical Physics \textbf{43}, 4273 (2002).

\bibitem{ref:SRFB} U. Schollw\"{o}ck, J. Richter, D. Farnell, and R. Bishop, Quantum Magnetism, Springer-Verlag, Berlin (2004).

\bibitem{ref:zheng}
S. B. Zheng, G. C. Guo, Phys. Rev. Lett. \textbf{85}, 2392 (2000).

\bibitem{ref:latorre}
G. Vidal, J. I. Latorre, E. Rico, and A. Kitaev, Phys. Rev.
Lett. \textbf{90}, 227902 (2003);

\bibitem{ref:SB} S. Bose, Phys. Rev. Lett. \textbf{91}, 207901 (2003).

\bibitem{ref:BMD70} E. Barouch, B. M. McCoy, and M. Dresden. Phys. Rev. A, \textbf{2}, 1075–1092 (1970).

\bibitem{ref:SS} A. Dutta and D. Sen,
Phys. Rev. B \textbf{67}, 094435 (2003); I. Peschel and V. J. Emery, Z. Phys. B \textbf{43}, 241 (1981); J. Kurmann, H. Thomas, and G. M\"{u}ller, Physica A \textbf{112}, 235 (1982); G. M\"{u}ller and R. E. Shrock, Phys. Rev. B \textbf{32}, 5845 (1985).

\bibitem{ref:brennen}
G. K. Brennen, Quant. Inf. Comput. \textbf{3}, 616 (2003).

\bibitem{ref:oliveira}
T. R. de Oliveira, G. Rigolin, M. C. de Oliveira, and E.
Miranda, Phys. Rev. Lett. \textbf{97}, 170401 (2006).

\bibitem{ref:montakhab2} A. Montakhab and A. Asadian
Phys. Rev. A \textbf{77}, 062322 (2008).

\bibitem{ref:LBS} P. J. Love, A. M. Brink, A. Y.  Smirnov et al. Quant. Inf. Process. \textbf{6}, 187 (2007).

\bibitem{ref:Scott} A. J. Scott Phys. Rev. A \textbf{69}, 052330 (2004).

\bibitem{ref:LSM} E. Lieb, T. Schultz, and D. Mattis, Ann. Physics \textbf{16}, 407 (1961).

\bibitem{ref:osborne}
T.J. Osborne and M. Nielsen, Phy. Rev. A \textbf{66}, 032110 (2002).

\bibitem{ref:BM71}
E. Barouch, and B. M. McCoy, Phys. Rev. A \textbf{3}, 786–804 (1971).

\bibitem{ref:MNB} M.N. Barber, \textit{in Phase Transitions and Critical Phenomena}, Academic, London, Vol. \textbf{8}, 146-259 (1983).

\bibitem{ref:McCoy} B. M. McCoy, Phys. Rev. A \textbf{3}, 786–804 (1971).

\bibitem{ref:PCF} A. D. Pasquale, G. Costantini, P. Facchi, G. Florio, S. Pascazio, K.Yuasa, Eur. Phys. J. Spec. Top. \textbf{160} 127 (2008).

\bibitem{ref:PF} A. D. Pasquale and P. Facchi
Phys. Rev. A \textbf{80}, 032102 (2009).

\bibitem{ref:JLEI} J. Larson and E. K. Irish, J. Phys. A: Math. Theor. \textbf{50}, 17 (2017).

\bibitem{ref:Pfuty} P. Pfeuty, Ann. Physics \textbf{57}, 79 (1970).

\bibitem{ref:MRBD} M. M. Rams and B. Damski
Phys. Rev. A \textbf{84}, 032324 (2011).

\bibitem{ref:CL} R. W. Cherng and L. S. Levitov
Phys. Rev. A \textbf{73}, 043614 (2006).

\bibitem{ref:MZPT} M. Zhong and P. Tong, J. Phys. A: Math. Theor. \textbf{43}, 505302 (2010).

\end{references}
\end{document}